%% file: main.tex
\documentclass[main]{ssh}
\usepackage[utf8]{inputenc}
\usepackage{amsmath,bm,amsfonts,gensymb}
\DeclareMathOperator*{\argminop}{arg\,min}

\myexternaldocument{SI}

\title{Imaging 3D Chemistry at 1 nm Resolution with Fused Multi-Modal Electron Tomography}

\author[1]{Jonathan~Schwartz}
\author[2]{Zichao~Wendy~Di}
\author[3]{Yi~Jiang}
\author[1]{Jason~Manassa}
\author[1,4]{Jacob~Pietryga}
\author[5]{Yiwen~Qian}
\author[5,11]{Min~Gee~Cho}
\author[6]{\\Jonathan~L.~Rowell}
\author[7]{Huihuo Zheng}
\author[8,9]{Richard~D.~Robinson}
\author[10]{Junsi~Gu}
\author[10]{Alexey~Kirilin}
\author[10]{Steve~Rozeveld}
\author[11]{Peter~Ercius}
\author[12]{\\Jeffrey~A.~Fessler}
\author[5,13]{Ting~Xu}
\author[5,11]{Mary~Scott}
\author[1,14,*]{Robert~Hovden}

\affil[1]{Department of Materials Science and Engineering, University of Michigan, Ann Arbor, MI}
\affil[2]{Mathematics and Computer Science Division, Argonne National Laboratory, Lemont, IL}
\affil[3]{Advanced Photon Source Facility, Argonne National Laboratory, Lemont, IL}
\affil[4]{Department of Material Science and Engineering, Northwestern University, Evanston, IL}
\affil[5]{Department of Materials Science and Engineering, University of California at Berkeley, Berkeley, CA}
\affil[6]{Department of Chemistry and Chemical Biology, Cornell University, Ithaca, NY}
\affil[7]{Argonne Leadership Computing Facility, Argonne National Laboratory, Lemont, IL}
\affil[8]{Department of Material Science and Engineering, Cornell University, Ithaca, NY}
\affil[9]{Kavli Institute at Cornell for Nanoscale Science, Cornell University, Ithaca, NY}
\affil[10]{Dow Chemical Co., Midland, MI}
\affil[11]{National Center for Electron Microscopy, Molecular Foundry,
Lawrence Berkeley National Laboratory, Berkeley, CA}
\affil[12]{Department of Electrical Engineering and Computer Science, University of Michigan, Ann Arbor, MI}
\affil[13]{Materials Science Division,
Lawrence Berkeley National Laboratory,
Berkeley, CA}
\affil[14]{Applied Physics Program, University of Michigan, Ann Arbor, MI}
\affil[*]{e-mail: hovden@umich.edu}
\date{\today}

\begin{document}

\abstract{Measuring the three-dimensional (3D) distribution of chemistry in nanoscale matter is a longstanding challenge for metrological science. The inelastic scattering events required for 3D chemical imaging are too rare, requiring high beam exposure that destroys the specimen before an experiment completes. Even larger doses are required to achieve high resolution. Thus, chemical mapping in 3D has been unachievable except at lower resolution with the most radiation-hard materials. Here, high-resolution 3D chemical imaging is achieved near or below one nanometer resolution in a Au-Fe$_3$O$_4$ metamaterial, Co$_3$O$_4$ - Mn$_3$O$_4$ core-shell nanocrystals, and ZnS-Cu$_{0.64}$S$_{0.36}$ nanomaterial using fused multi-modal electron tomography. Multi-modal data fusion enables high-resolution chemical tomography often with 99\% less dose by linking information encoded within both elastic (HAADF) and inelastic (EDX / EELS) signals. Now sub-nanometer 3D resolution of chemistry is measurable for a broad class of geometrically and compositionally complex materials.


}

\titlemaker

\section*{Introduction}

Knowing the complete chemical arrangement of matter in all dimensions is fundamental to engineering novel nanomaterials \cite{gang2022xrayPtychoXRF}. Although electron tomography provides comprehensive 3D structure at resolutions below 1 nm using elastic scattering signals~\cite{scott2012electron,yang2021determining,levin2016nanomaterial}, chemical tomography obtained from inelastic scattering remains largely out of reach. Several demonstrations of chemical tomography using electron energy loss or x-ray energy spectroscopy (EELS / EDX) accompanied the introduction of scanning transmission electron microscope (STEM) tomography and provide a milestone for 3D imaging~\cite{mobus2001eftemTomo,midgley2001zConstrastTomo,midgley2013plasmonTomo,lepinay2013transistorTomo}. However, chemical tomography from core-excitation spectroscopy demands high electron doses that almost always exceed the specimen limits (e.g.,~$>10^7$~e/Å$^2$)~\cite{cueva2012csi,hart2017direct,midgley2017chemTomoReview}. If attempting chemical tomography, researchers must sacrifice resolution by collecting few specimen projections (e.g., 5-10) and constrain the total dose (e.g., $<10^6$~e/Å$^2$). Consequently, 3D resolution is penalized from undersampling and noisy chemical maps~\cite{crowther1970resolutionCriterion}. Therefore, a paradigm shift is necessary for high-resolution chemical tomography.


We show achieving high-resolution 3D chemistry at lower dose requires fusing both elastic and inelastic scattering signals. Typically these detector signals are analyzed separately and correlated \cite{goris2014edxTomoGalvanic,miao2019meteorite,ruoqian2019edxTomoLithium}. However, correlative imaging disregards shared  but also complementary information between structure and chemistry and misses opportunities to recover useful information \cite{su2010correlativeImg}. Data fusion, popularized in satellite imaging, goes further than correlation by linking separate signal modalities to reconstruct new information and improve measurement accuracy~\cite{hall1997introduction,lahat2015datafusionreview,wendy2017jointXray}. Recent developments in multi-modal data fusion paved new opportunities for high-resolution chemical imaging by substantially reducing the dose requirements to successfully acquire an atomic-resolution map~\cite{schwartz2022emMM}. In alignment with the principles of fused multi-modal electron microscopy, we extend its algorithmic framework into the third dimension.

\begin{figure*}[ht!]
    \centering
    \includegraphics[width=\linewidth]{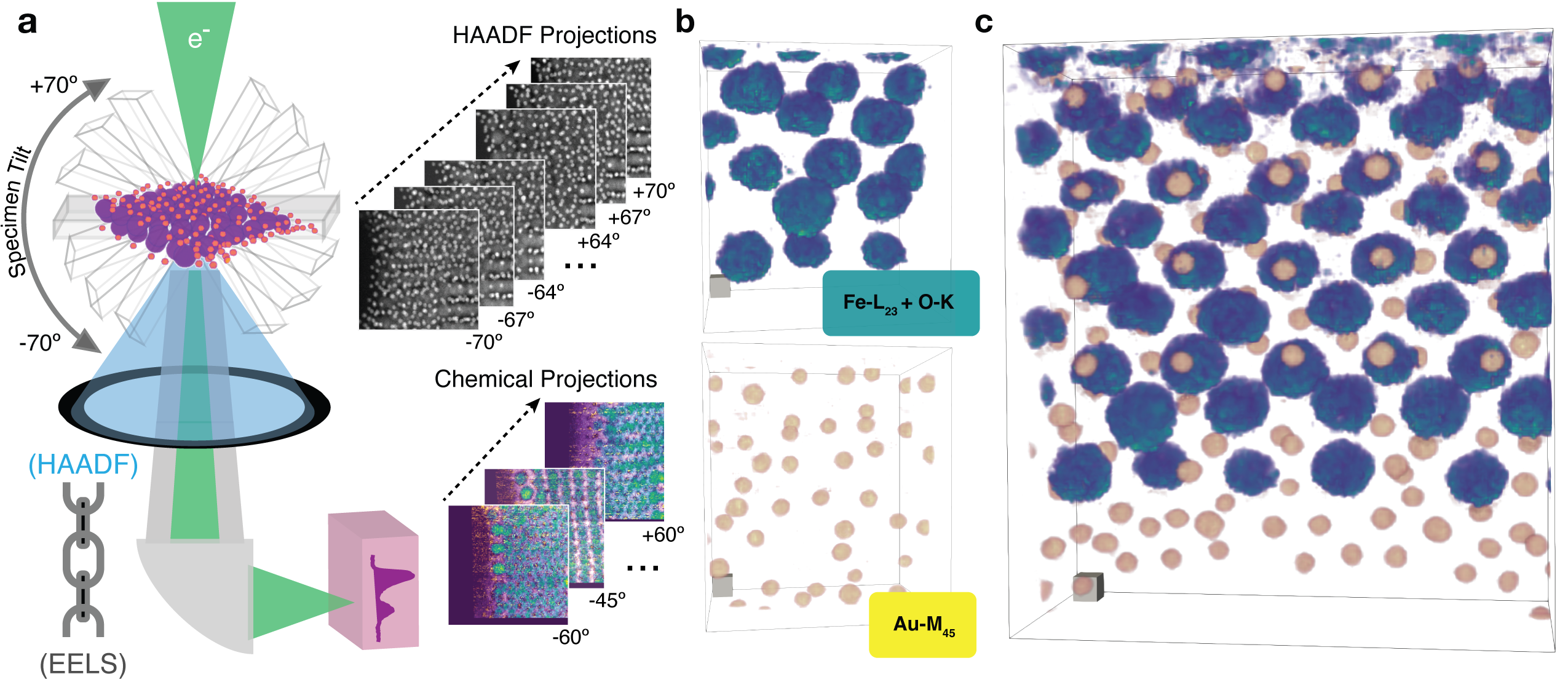}
    \caption{\textbf{Nanoscale recovery of Au-Fe$_3$O$_4$ nanoparticle superlattice.} \textbf{a} Schematic highlighting the linked HAADF and EELS modalities for chemical tomography. HAADF projection images are collected at every tilt increment while core-loss EELS spectra are sparsely acquired every few tilts.  \textbf{b} The fused multi-modal reconstruction for the specimen's Fe L$_{2,3}$ (turquoise), O-K (turquoise), and gold M$_{4,5}$ edge (yellow). \textbf{c} Chemical overlay of the superlattice nanoparticles over the entire 115 nm field of view. Scale cubes, 5 nm$^3$.}
    \label{fig::tomo_eels}
\end{figure*}

Here we introduce fused multi-modal electron tomography, which offers high signal-to-noise (SNR) and high-resolution recovery of material chemistry in 3D by linking information encoded within both elastic high-angle annular dark field (HAADF) and inelastic (EDX / EELS) scattering signals.  Multi-modal electron tomography reconstructs the volumetric chemical structure of specimens by solving a 3-term inverse problem that fuses signals from multiple detectors.  This framework enables new sampling strategies that minimize dose by measuring a high number of HAADF projections alongside far fewer chemical projections---dose reductions of one-hundred fold are readily achieved. Although the chemical structure is severely underdetermined, fusing the two modalities fills in missing information, notably improving resolution and reconstruction quality. Our approach demonstrates that researchers can measure 3D chemistry at 1 nm resolution using electron doses as low as $10^4$ e/Å$^{2}$ and as few as 9 spectroscopic maps while remaining consistent with original measurements. Multi-modal tomography is validated across multiple material systems, including Au-Fe$_3$O$_4$ superlattice clusters, core-shell Co$_3$O$_4$-Mn$_3$O$_4$~\cite{hwan2020comno}, ZnS-Cu$_{0.64}$S$_{0.36}$ heterostructures~\cite{ha2014CuZnNP}, Cu-SiC nanoparticles and a range of simulated specimens. By fusing modalities, chemical tomography is now possible at sub-nanometer resolution for a wider class of material systems. 

\section*{Results}
\subsection*{Principles of Fused Multi-Modal Electron Tomography}

High-resolution 3D chemical imaging is achieved using the multi-modal electron tomography framework illustrated in Fig.~\ref{fig::tomo_eels}a for a binary Au-Fe$_3$O$_4$ nanoparticle superlattice within a carbon-based matrix. In multi-modal electron tomography, projections of the specimen structure are measured from a HAADF detector and the specimen chemistry is extracted from spectroscopy (EELS or EDX). These two detector modalities are fused during the reconstruction process to provide the complete 3D chemical distribution of a specimen at high resolution and SNR. Figure~\ref{fig::tomo_eels}b shows the 3D reconstruction of each individual chemistry: larger 10.2~$\pm$~1.1~nm Fe nanoparticles (blue) and smaller Au 3.9~$\pm$~0.4 nm nanoparticles (orange). Both chemistries are visualized simultaneously in Fig~\ref{fig::tomo_eels}c to show the self-organization of the chemical superlattice. The light-element, carbon matrix is shown in Supplemental Figure~\ref{sfig:AuFeO_withCarbon}.

In multi-modal tomography, the number of structural HAADF projections usually exceeds the chemical projections. In this first demonstration, only 9 chemical maps ($\Delta \theta = 15^{\circ}$) are measured from the Fe-L$_{2,3}$ and Au-M$_{4,5}$ core-excitation edges in an EELS spectrum whereas 47 HAADF images ($\Delta \theta = 3^{\circ}$) are collected over a $\pm 70^{\circ}$ specimen tilt range. Linking both modalities into the reconstruction enables a clear distinction between Fe$_3$O$_4$ and Au nanoparticles at high resolution from just a few EELS maps and a total electron dose of $~5\times 10^{5}$ e/Å$^{2}$---roughly two orders of magnitude lower total electron dose than an equivalent conventional approach.


Fused multi-modal electron tomography reconstructs three-dimensional chemical models by solving an optimization problem seeking a solution that strongly agrees with (1) the HAADF modality containing high SNR, (2) the chemically sensitive spectroscopic modality (EELS and / or EDX), and (3) encourages sparsity in the gradient domain producing solutions with reduced spatial variation. The overall optimization function is as follows:
\begin{align}
\label{eq:costFunc}
     &\argminop_{\bm{x}_i \geq 0} \quad \frac{\lambda_1}{2} \Big\| \bm{A}_h \sum_{i} (Z_i\bm{x}_{i})^\gamma - \bm{b}_{H} \Big \|_2^2 +
     \nonumber\\
     \lambda_2 \sum_{i} & \Big(\bm{1}^T \bm{A}_c \bm{x}_i - \bm{b}_{i}^T \log(\bm{A}_c \bm{x}_i + \varepsilon) \Big) + \lambda_3 \sum_{i} \|\bm{x}_i\|_{\mathrm{TV}},
\end{align} 
$\bm{x}_i$ is the reconstructed 3D chemical distributions for element $i$, $\bm{b}_i$ is the measured 2D chemical maps for element $i$, $\bm{b}_H$ is the measured HAADF micrographs, $\bm{A}_h$ and $\bm{A}_c$ are forward projection operators for HAADF and chemical modalities, $\lambda$ are regularization parameters, $\varepsilon$ herein prevents log(0) issues but can also account for background, the $\log$ is applied element-wise to its arguments, superscript $T$ denotes vector transpose, and $\bm{1}$ denotes the vector of $N^{\mathrm{proj}}_{\mathrm{chem}} n_y n_{i}$ ones, where $n_y$ is the number of pixels, $n_{i}$ is the number of elements present, and $N^{\mathrm{proj}}_{\mathrm{chem}}$ is the number of projections for the chemical modality. Pseudo-code for numerical implementation is provided in the Supplemental Materials.

\begin{figure*}[ht!]
    \centering
    \includegraphics[width=\linewidth]{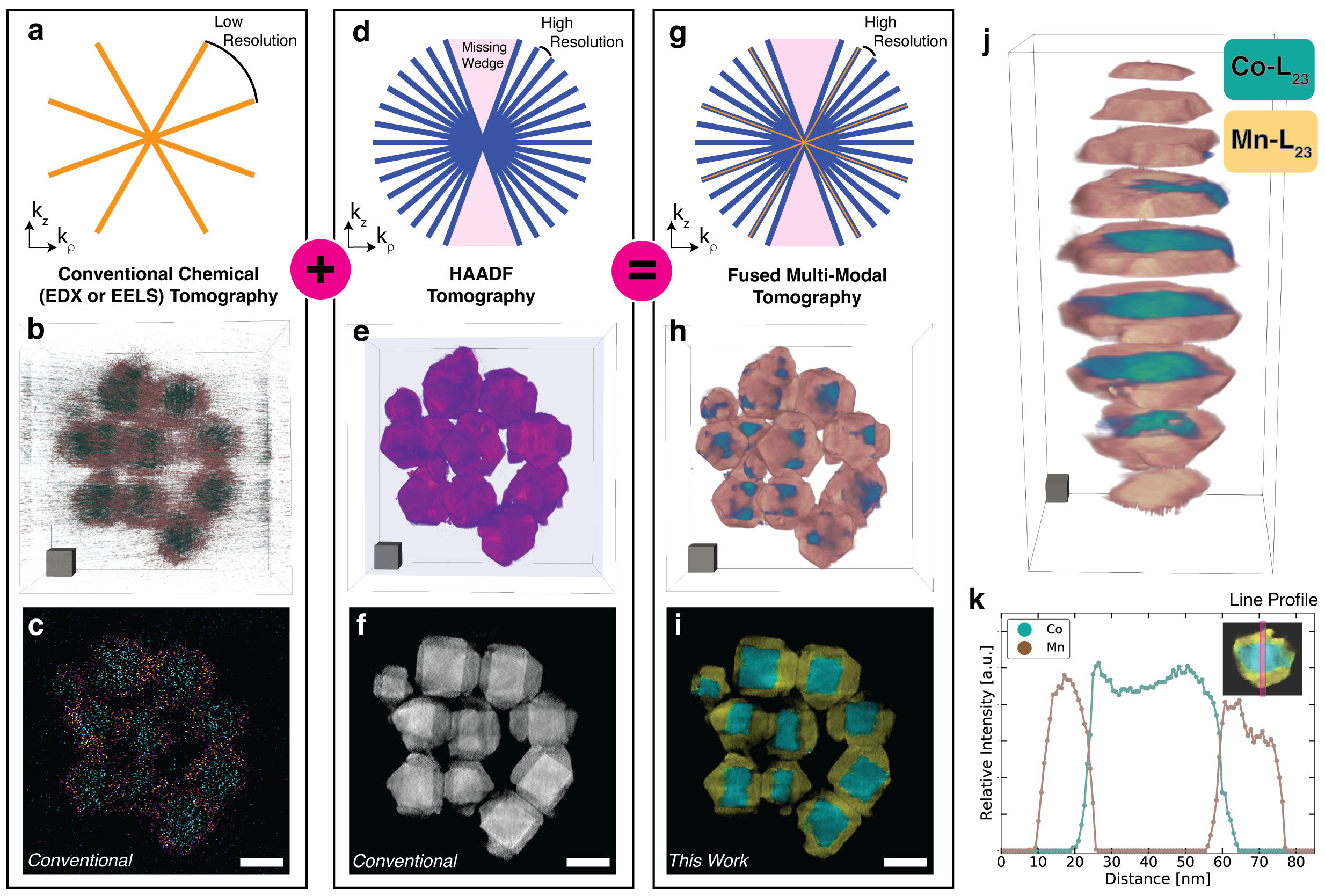}
    \caption{\textbf{Nanoscale recovery of Co$_3$O$_4$-Mn$_3$O$_4$ core-shell nanoparticles.} \textbf{a-c}  Raw EELS reconstruction for the Co (blue-green) and Mn (orange) L$_{2,3}$ core-loss edges.\textbf{d-f} The HAADF tomogram of Co$_3$O$_4$-Mn$_3$O$_4$ nanoparticle tracks the structure of the specimen but fails to describe materials chemistry in 3D. \textbf{g-i} The fused multi-modal reconstruction.  Scale cubes, 25 nm$^3$. \textbf{a,d,g} Representation in Fourier space of the projections used to reconstruct the tomograms. \textbf{j} Fused multi-modal tomogram of a single  Co$_3$O$_4$-Mn$_3$O$_4$ nanoparticle. Scale cube, 10 nm$^3$. \textbf{e} A line profile showing the average intensity across the diameter of the particle.}
    \label{fig::tomo_CoMnO}
\end{figure*}

The three terms in Equation \ref{eq:costFunc} define our fused multi-modal framework designed to surpass traditional limits for chemical tomography. First, we assume a forward model where the simultaneous HAADF is a linear combination of the reconstructed 3D elemental distributions ($\bm{x}_i^\gamma$ where $\gamma \in$ [1.4, 2]). The incoherent linear imaging approximation for elastic scattering scales with atomic number as $Z_i^\gamma$, where experimentally $\gamma$ is typically around 1.7~\cite{hartel1996conditions,krivanek2010atom,hovden2012singleAtomImaging}. This $\gamma$ is bounded between 4/3 as described by Lenz-Wentzel expressions for electrons passing through a screened coulombic potential and 2 for Rutherford scattering from bare nuclear potentials~\cite{crewe1970coloumb, langmore1973eScatter}. Second, we ensure the recovered 3D distributions maintain a high degree of data fidelity with the initial measurements by using the log-likelihood for spectroscopic measurements dominated by low-count Poisson statistics~\cite{wendy2017jointXray,odstrcil2018iterativeLS}. In a higher count regime, this term can be substituted with a least-squares discrepancy ($\| \bm{Ax} - \bm{b} \|_2^2$)~\cite{csiszar1991poissonLS}. Lastly, we include channel-wise isotropic total variation (TV) regularization to enforce a sparse gradient magnitude, which reduces noise by promoting image smoothness while preserving sharp features~\cite{osher1992tv}. This sparsity constraint, popularized by the field of compressed sensing (CS), is a powerful yet modest prior for recovering structured data~\cite{donoho2006CS,candes2006CS}. When solving Equation~\ref{eq:costFunc}, each of these three terms should be weighted appropriately by determining coefficients ($\lambda$) that balance their contributions. Ultimately, optimization of all three terms is necessary for accurate recovery (Supplementary~Fig.~\ref{sfig:CoCuOComponentCost}-\ref{sfig:CoNiOComponentCost}).
 
The improvement in reconstruction quality with fused multi-modal chemical tomography (Fig.~\ref{fig::tomo_CoMnO}i) is dramatic when compared to traditional chemical tomography (Fig.~\ref{fig::tomo_CoMnO}c).


\subsection*{3D Chemistry at High-Resolution, Low-Dose}

 In tomography, 3D resolution is described by the Crowther criterion, which states resolution is limited by the object size and the number of specimen projections measured~\cite{klug1972crowtherInfoTheory} -- higher resolution requires more projections~\cite{reed2021throughFocalSim}. For traditional chemical tomography, few chemical projections are collected and the Crowther relation devastates resolution in 3D. This limitation occurs from the high-dose requirements of chemical mapping (i.e., EDX, EELS) where only a few projections can be collected before radiation damage alters the specimen structure.

\begin{figure}[ht!]
    \centering
    \includegraphics[width=\linewidth]{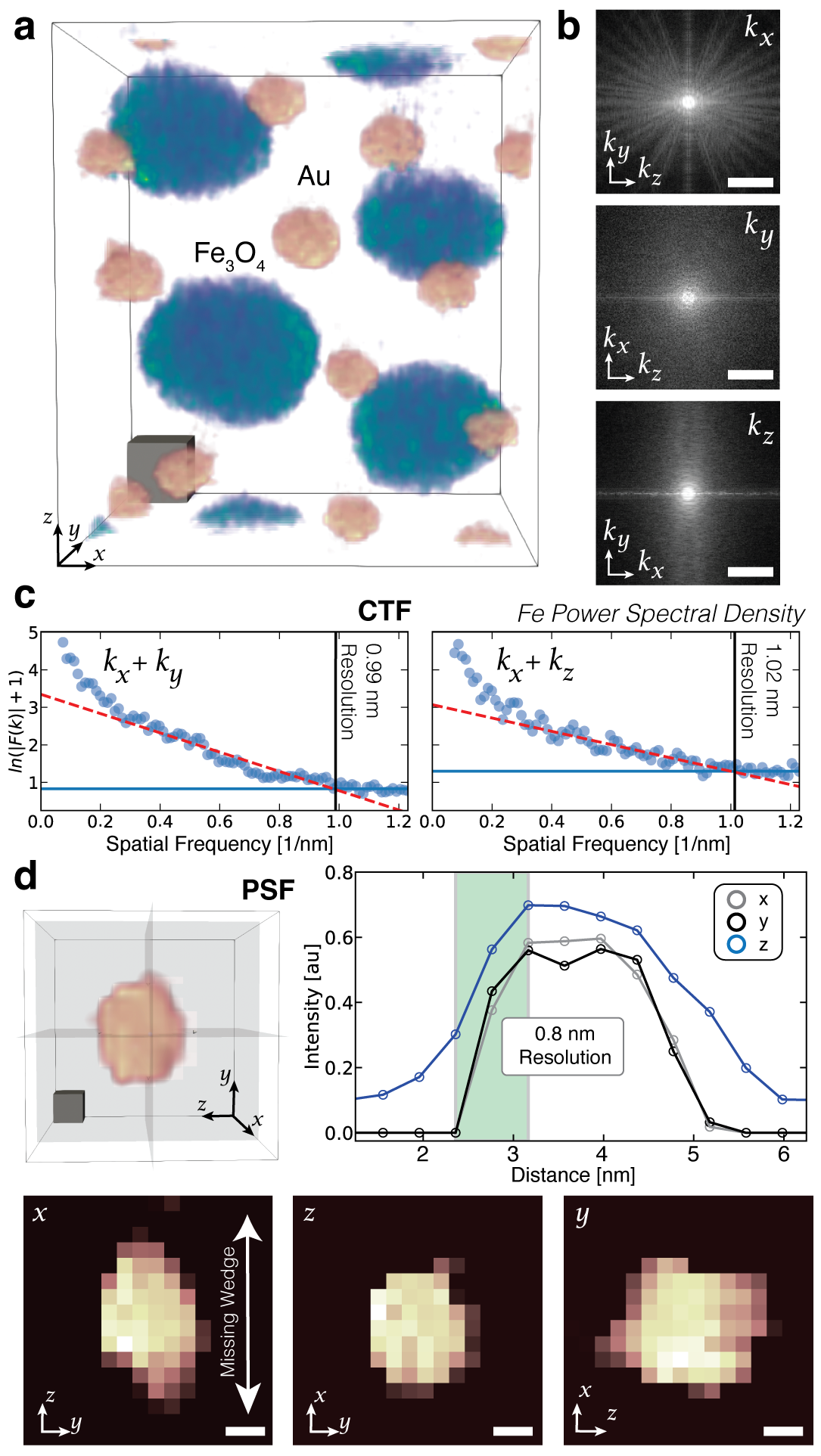}
    \caption{\textbf{Resolution Analysis of Au-Fe$_3$O$_4$ superlattice nanoparticles.} \textbf{a} Fused EELS tomograms of Au-Fe$_3$O$_4$ nanoparticles.  Scale cube, 2 nm$^3$. \textbf{b} Power spectral density of the Fe reconstruction along the principal axial directions shown on the right. Scale bar, 0.5 nm$^{-1}$. \textbf{c} Power spectral density profiles for $k_x$-$k_y$ and $k_x$-$k_z$ directions. \textbf{d} Line  profiles of a 2.5 nm Au nanoparticle gives a resolution of 0.8 nm, 0.8 nm, and 1.1 nm along the x, y, and z directions. }
    \label{fig::tomo_resolution}
\end{figure}
 
Figure~\ref{fig::tomo_CoMnO} shows how specimen projections from each modality are superimposed as planes of information in Fourier space. Chemical tomography is sparsely sampled in Fourier space (Fig.~\ref{fig::tomo_CoMnO}a), which results in a tomographic reconstruction containing artifacts and low SNR (Fig.~\ref{fig::tomo_CoMnO}b,c). Despite the poor quality, traditional chemical tomography tracks the chemical distribution, and the Mn shell  (orange) can be seen surrounding the Co core (blue-green). In contrast, elastically scattered electrons collected by the HAADF detector provide high signals at lower doses and allow many projections to be collected---in practice, HAADF sampling is five to ten times more finely spaced than chemical (Fig.~\ref{fig::tomo_CoMnO}d)~\cite{hovden2012singleAtomImaging}. The dose required for a single HAADF projection is 10$^2$-10$^3$ times lower than a chemical projection acquired using core-energy loss spectroscopy. Thus, it is favorable to acquire more HAADF images and achieve higher resolution. Although HAADF tomography permits high-resolution and high-SNR reconstructions of structure, it lacks chemical specificity. This is seen in Figure~\ref{fig::tomo_CoMnO}e,f where the structure is well defined with low noise but the Co and Mn regions are not identifiable.


Exploiting shared information in both modalities, multimodal tomography achieves a chemical resolution in 3D comparable to high-resolution HAADF reconstructions. Although few chemical measurements pose a severely underdetermined problem, fusing with the HAADF modality fills in missing chemical information. This is reflected in Figure~\ref{fig::tomo_CoMnO}g where many HAADF projections (e.g.,~50-180) are measured while far fewer chemical projections (e.g.,~5-15) are intermittently measured.  In this reconstruction, 9 EELS maps and 45 HAADF projections (50-200 mrad detector inner and outer semi-angles) were collected over a $\pm$60$^\circ$ tilt range using a 2.4 Å probe with a 24.3 nm depth of focus (300 keV acceleration voltage, 10 mrad convergence angle). High-resolution of 3D chemistry is visible in the the core shell Co$_3$O$_4$-Mn$_{3}$O$_{4}$ using multi-modal tomography in Figure~\ref{fig::tomo_CoMnO}h,i. 

Fused multi-modal electron tomography provides unique insight for studying heterostructured nanocrystals with unprecedented geometries. In the case of Co$_3$O$_4$ - Mn$_3$O$_4$ nanocrystals, the manganese oxide shell is divided into several ordered grains that grow on each surface plane for the cobalt oxide nanocube core \cite{hwan2020comno}. However the core and shell interface can vary per plane driven by the growth interplay between  strain and surface energy, resulting in the formation of grain boundaries \cite{shklyaev2005strainNP}. The complete 3D distribution of Co and Mn at the surface and interface is difficult to discern with 2D projected EELS maps or HAADF reconstructions. Fortunately, the fused chemical distributions reveals surface coverage of the shell grains and cross-sections quantify the shell thickness and interface chemistry. To further demonstrate, fused multi-modal EELS tomography was used to discern between ZnS and Cu$_{0.64}$S$_{0.36}$ phases (Supplementary Fig.~\ref{sfig:ZnSCuS_abstrat}) in a heterostructured nanocrystal~\cite{ha2014CuZnNP} and EDX tomography to identify Cu nanoparticles embedded in SiC catalysts (Supplementary Fig.~\ref{sfig:CuSiC_EDX}).


Data fusion eliminates noticeable noise in the final 3D chemical reconstruction without a loss of resolution. This noise reduction accompanies a dose reduction of roughly one-hundred fold. Linking the chemical projections to the high SNR HAADF signals dose-efficiently boosts the chemical specificity. To illustrate, in Figure~\ref{fig::tomo_CoMnO}, matching the resolution of fused multi-modal chemical tomography using traditional methods would require 45 EELS maps---a five-fold dose increase. However, the SNR of each chemical projection would still fall short (Supplementary~Fig.~\ref{sfig:RawEELSCoMnOTilts}) and requires roughly twenty-times additional dose. In total, multi-modal chemical tomography performs well at one-hundredth the dose requirement of traditional methods.

\subsection*{Sub-nanometer Chemical Resolution in 3D}

3D resolution of the chemical distribution in Au-Fe$_3$O$_4$ nanoparticle superlatice (Fig.~\ref{fig::tomo_resolution}a) is demonstrated at or below 1 nm using multi-modal tomography. The achieved resolution is quantified in real and reciprocal space.  In real space, the resolution limit is verified by visually inspecting a single 3 nm Au nanoparticle (Fig.~\ref{fig::tomo_resolution}d). The edge sharpness between the reconstructed nanoparticle and vacuum is visibly less than 1 nm. From line profiles, the half pitch resolution is 0.8 nm~$\times$~0.8 nm~$\times$ 1.1 nm along the \textit{x}, \textit{y}, and \textit{z} directions respectively. Along optimal directions (\textit{x}, \textit{y}) the resolution is comparable to the Nyquist frequency (8.05 Å). The real-space resolution is consistent with reciprocal space estimates of the cutoff frequency at which the signal drops to the noise floor~\cite{gang2022xrayPtychoXRF}. Figure~\ref{fig::tomo_resolution}b highlights power spectral density variations projected on three orthogonal planes. Measured power spectral density along the $k_x$-$k_y$ and $k_x$-$k_z$ directions shows information transfer roughly occurring at 0.99 nm and 1.02 nm  respectively  (Fig.~\ref{fig::tomo_resolution}c). These directions conservatively represent the 3D resolution from an average of the high-resolution and low-resolution (z-axis) directions. This 3D chemical resolution nearly matches the 3D HAADF resolution 1.00 nm, 1.01 nm in  Figure~\ref{fig::tomo_resolution} (Supplementary Fig.~\ref{sfig::haadf_resolution}). For fused multi-modal chemical tomography, the HAADF 3D resolution provides a new upper bound to the highest obtainable 3D chemical resolution. A reduction of resolution along the \textit{z}-axis is expected from the incomplete tilt range that creates a missing wedge of information in Fourier space~\cite{leary2013cset}. Here, we observe approximately a 25\% reduction in resolution along the missing wedge direction of the multi-modal chemical reconstruction.


\subsection*{Influence of Sampling}

\begin{figure}[ht!]
    \centering
    \includegraphics[width=0.95\linewidth]{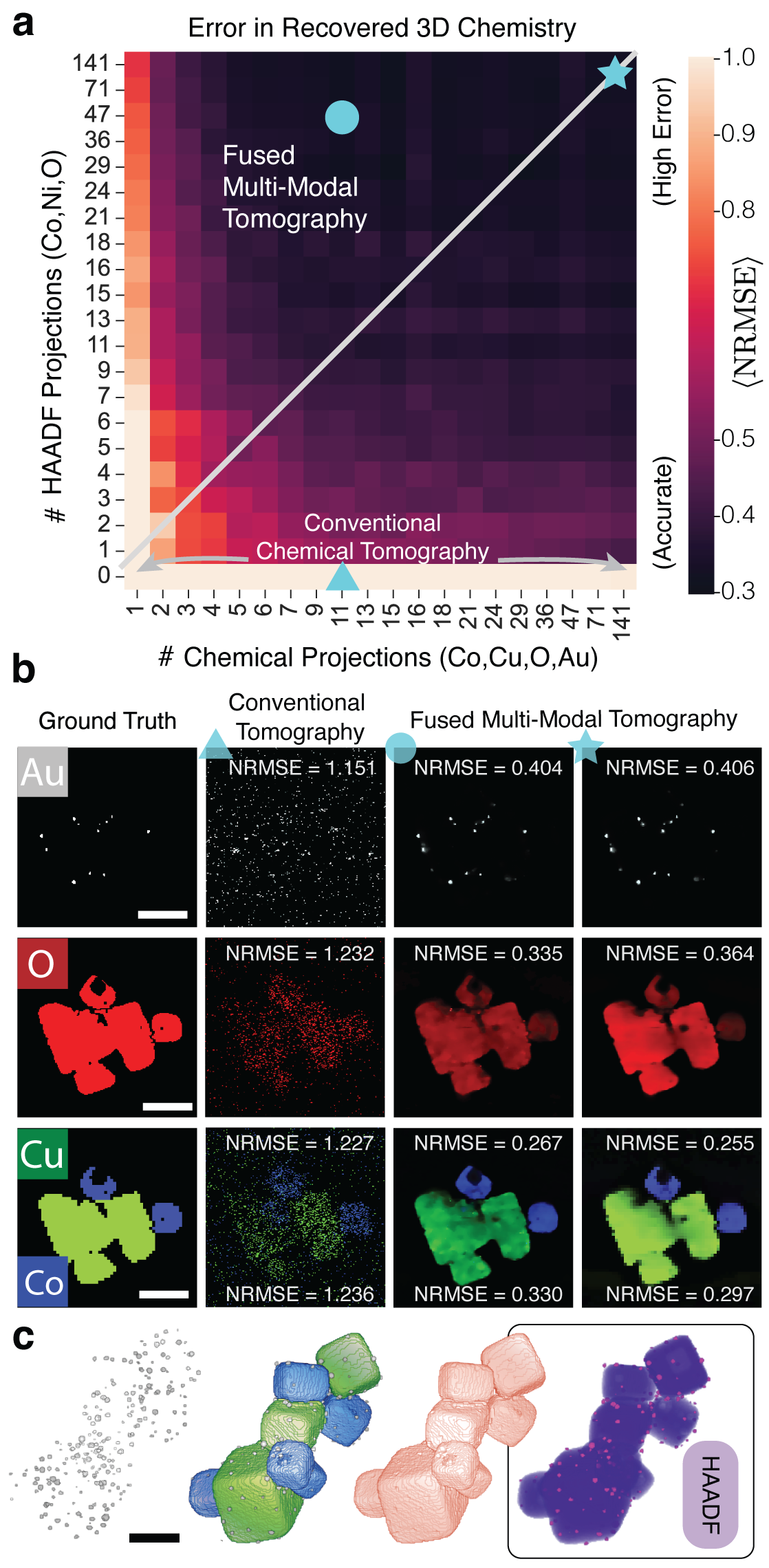}
    \caption{\textbf{Estimating Sampling Requirements for Accurate Recovery with Synthetic CoO/CuO Nanocubes.} \textbf{a} An NRMSE map representing the reconstruction error as a function of the number of HAADF and chemical tilts. Brighter pixels denote results containing incorrect reconstructions from the ground truth. \textbf{b} Visualization of three points corresponding to conventional chemical tomography (reconstruction without the HAADF modality), and low or high-dose fused multi-modal electron tomography. \textbf{c} The 3D models used for generating synthetic chemical and ADF projections. Scale bar, 75 nm.}
    \label{fig::tomo_sim}
\end{figure}

Electron tomography simulations show a 3-5 fold improvement in the normalized root mean square error $\big(\langle \text{NRMSE} \rangle\big)$ averaged across all elements when multi-modal tomography is used over conventional chemical tomography. In Figure~\ref{fig::tomo_sim} synthetic gold decorated CoO / CuO nanocubes inspired by real experimental data~\cite{padget2017auSTO} provide a ground truth comparison to assess the accuracy of fused multi-modal tomography. Simulated projection images are generated from a simple linear incoherent imaging model of the 3D chemical composition with added Poisson noise (See Methods). The specimen tilt range is limited to $\pm 70 ^{\circ}$ to better match typical experimental conditions. The advantages of multi-modal tomography are clearly visible in the 2D slices (Fig.~\ref{fig::tomo_sim}b) taken from 3D reconstructions obtained by conventional chemical tomography $\big(\langle\text{NRMSE}\rangle = 1.301 \big)$ and fused multi-modal tomography $\big(\langle\text{NRMSE}\rangle = 0.33 \big)$. For all chemistries (Au, O, Cu, Co,) fused multi-modal tomography is more consistent with the ground truth with higher resolution and reduced noise.


For any number of chemical projections acquired, we see a notable reduction in NRMSE when HAADF projections are integrated into the chemical reconstruction. Figure~\ref{fig::tomo_sim} shows the improved fused multi-modal reconstruction accuracy across a wide range of HAADF and chemical projections for the gold-decorated CoO / CuO nanocubes. The reconstruction error (average NRMSE) across most of the multi-modal parameter space is less than 0.40 compared to values around 1.2 for conventional tomography. Pixel values on the diagram (Fig.~\ref{fig::tomo_sim}a) represent the average NRMSE across all of the elements. This NRMSE map shows data fusion strongly benefits by increasing the HAADF information available. It requires substantially less dose to increase the HAADF projections (i.e. moving vertically on the map) compared to increasing the chemical projections (i.e. moving horizontally on the map).

Conventional chemical tomography does not use HAADF projections (bottom row, Fig. \ref{fig::tomo_sim}a) resulting in an average reconstruction error larger than the entire multi-modal regime. In practice fused multi-modal tomography is performed in the regime with equal or more HAADF projections than chemical (i.e. top-left triangle).  Multi-modal tomography also performs well when the chemical projections exceed the number of HAADF projections, however, this is not practical since HAADF signals can be acquired simultaneously with EDX and EELS. Similar trends are observed in a second large-scale simulation performed on a synthetic composite structure composed of transition metal CoO nanoparticles embedded in a NiO support (Supplementary~Fig.~\ref{sfig:CoNiOphaseMap}).

\section*{Discussion}

While this paper highlights the advantages of fused multi-modal electron tomography, the technique is not a simple black-box solution. Step sizes for convergence and weights on the terms in the cost function (Eq.~\ref{eq:costFunc}) must be reasonably selected. Standard spectroscopic pre-processing methods become ever more critical in combination with multi-modal fusion. Improper background subtraction of EELS spectra~\cite{muller2008atomicEELS} or overlapping characteristic X-ray peaks that normally cause inaccurate stoichiometric quantification also reduces the accuracy of fused multi-modal tomography. Thick specimens with dimensions that far exceed the mean free path of the electron can produce inversion contrast that will cause electron tomography to fail~\cite{ercius2006ibfTomo}---also causing failure for multi-modal electron tomography (Supplementary~Fig.~\ref{sfig:TiO2}). As shown for 2D fused multi-modal electron microscopy~\cite{schwartz2022emMM}, fused multi-modal tomography works best when elements have discernible contributions to the HAADF contrast and all chemical elements have been imaged. Multi-modal tomography leverages compressed sensing (e.g. TV min.) which assumes incoherence (i.e., a high level of dissimilarity) between the sensing and sparsifying transform~ \cite{candes2006incoherenceCS, lustig2007sparseMRI, schwartz2019destripe}---although this assumption typically holds as demonstrated for the datasets presented herein.




\section*{Conclusion}

In summary, we present fused multi-modal electron tomography that enables chemically-sensitive 3D reconstruction of matter with nanometer resolution at high SNR. Researchers no longer must choose between measuring 3D structure without chemical detail or characterizing chemistry along a single viewing direction. By linking signals from elastic (HAADF) and inelastic (EDX / EELS) scattering processes, the traditional dose limits of chemical tomography are substantially surpassed. In some cases, a one-hundred fold reduction in dose is estimated. To demonstrate, the complete volumetric density of each chemistry was mapped in several systems including Au-Fe$_3$O$_4$, Co$_3$O$_4$-Mn$_3$O$_4$, ZnS-Cu$_{0.64}$S$_{0.36}$, and Cu-SiC nanomaterials. In both synthetic and experimental datasets, fused multi-modal electron tomography shows substantial advantages in the accuracy of 3D chemical imaging. This approach enables chemical tomography of a wide range of previously inaccessible materials with moderate radiation sensitivity. Fused multi-modal electron tomography opens up new understanding of geometrically and compositionally complex materials.

Here, fused multi-modal tomography used commonly available STEM detectors (HAADF, EDX, and EELS), however, this approach can be extended to other modalities in development---including pixel-array detectors \cite{kayla2016empad}, annular bright field \cite{findlay2010abf}, ptychography \cite{jiang2018ptycho}, low-loss EELS \cite{hachtel2019vibrationalEELS}, etc. One can imagine a future wherein all scattered and emitted signals in an electron microscope are collected and fused for maximally efficient characterization of matter in all dimensions.





\input{appendix.tex}

\beginReferences{}

\section*{Acknowledgements}
R.H. and J.S. acknowledge support from the Dow Chemical Company. This work made use of the Michigan Center for Materials Characterization (MC2) and Molecular Foundry, Lawrence Berkeley National Laboratory. The Molecular Foundry was supported by the Office of Science, Office of Basic Energy Sciences, of the U.S. Department of Energy under Contract No. DE-AC02-05CH11231. The authors thank Tao Ma and Bobby Kerns for their assistance at MC2. R.D.R. acknowledges support from NSF under grant DMR-1809429. This research used the Oak Ridge Leadership Computing Facility at the Oak Ridge National Laboratory and Argonne Leadership Computing Facility at Argonne National Laboratory, which is supported by the Office of Science of the U.S. Department of Energy under Contract No. DE-AC05-00OR22725 and DE-AC02-06CH11357.

\section*{Author Contributions}
J.S., Y.J., and R.H. conceived the idea. J.S.,  J.P., and R.H. implemented the multi-modal reconstruction algorithms. J.F. and Z.W.D. validated algorithms and provided mathematical rigor. J.S., J.M., H.Z., and R.H. conducted the chemical tomography simulations. J.S. conducted the EELS and EDX tomography experiments designed by J.S., R.H., M.S., P.E.. Y.Q. and T.X. synthesized the Au-Fe$_3$O$_4$ nanoparticles and provided data interpretation. M.C. and M.S. synthesized the CoMnO nanoparticles and provided data interpretation. J.R. and R.R. synthesized the ZnS-CuS nanoparticles and provided data interpretation. S.R. and A.K. provided the Cu-SiC nanoparticles and data interpretation. J.G. provided the C-TiO$_2$ nanoparticles and data interpretation. J.S. and R.H. wrote the manuscript. All authors reviewed and edited the manuscript.

\section*{Ethics Declarations}
\subsection*{Competing Interests}
The authors declare no competing interests.

\end{document}

%% file: appendix.tex
\section*{Methods}

\subsection*{Specimen Synthesis and Preparation}

\textit{Au-Fe$_3$O$_4$ Superlattice Nanoparticles.} Syntheses of 3.9 nm Au NPs~\cite{peng2008auForCOoxidation} and 10.2 nm Fe$_3$O$_4$ NPs~\cite{park2004nanoCrystals} were carried out under nitrogen atmosphere using standard Schlenk line techniques according to literature methods. Polystyrene-based ligands were attached to the NP surface through a ligand exchange process as reported before~\cite{ye2015nanocrystals}. Thiol-terminated PS (PS-SH) was used as the polymeric ligand for Au NPs and was synthesized using Radical Addition Fragmentation Transfer (RAFT) polymerization and end-functionalized by aminolysis. Amine-terminated polystyrene was used as the polymeric ligand for Fe$_3$O$_4$ NPs and was synthesized using atom transfer radical polymerization and end-group modification~\cite{xiong2016coPolymer}. Binary superlattice of Au and Fe$_3$O$_4$ NPs was prepared by nanoparticle co-crystallization at water-air interface. A toluene solution containing two types of NPs with concentration ratio of 2:1 was drop-cast onto the water surface in a Teflon well and slowly dried overnight. The binary nanoparticle film was transferred onto a 200-mesh carbon TEM substrate and further dried in vacuum oven for 6 hours to remove residual solvent. 

\textit{Co$_3$O$_4$ nanocubes.} A mixture of 0.37 g of cobalt(II) perchlorate (Aldrich) and 2.7 g of oleylamine (Acros) in 15 mL of 1-octanol (Aldrich) was heated to 120 $\degree$C under air and aged for 2 hr. During the heating, 0.7 mL of distilled water was added before the temperature reaches 120 $\degree$C. After the reaction, an excess amount of acetone and ethanol was added and Co$_3$O$_4$ nanocubes were retrieved by centrifugation.

\textit{Core-Shell Co$_3$O$_4$-Mn$_3$O$_4$ Nanoparticles.} An organic/aqueous suspension was prepared by adding 0.080 g of Co$_3$O$_4$ nanocubes into a mixture of oleylamine (5 mmol), oleic acid (0.5 mmol), formic acid (3.15 mmol, Aldrich), and 15 mL of xylenes (Aldrich). The as-prepared suspension was heated to ~40 $\degree$C under air and aged for three hours with magnetic stirring. And then, 0.7 mL of ~0.7 M aqueous solution of manganese (II) chloride tetrahydrate was rapidly injected into the suspension at 90 $\degree$C and aged for 1.5 h under air. After the reaction, the nanocrystals were washed with hexane/ethanol and retrieved by centrifugation. The final product was prepared with three iterations of this process.

\textit{ZnS – Cu$_{0.64}$S$_{0.36}$ Nanocrystals.} Synthesis of the ZnS – Cu$_{0.64}$S$_{0.36}$ Heterostructured NPs was performed as described by literature using
typical air and water free synthetic techniques \cite{ha2014CuZnNP}. Cu$_{1.81}$S (roxbyite) nanocrystals are synthesized by
first dissolving CuCl$_{2}$ $\cdot$ 2H$_{2}$O in oleylamine (OLAM) at 200 $\degree$C after thoroughly degassing the solution at
high temperature. Tert-butyl-disulfide is then injected at 180 $\degree$C and the reaction continues at this
temperature for 40 minutes. After cooling to room temperature, the NPs are washed with hexanes and
acetone then dried in a vacuum desiccator.
The roxbyite NPs are then subjected to cation exchange as described previously \cite{ha2014CuZnNP} Briefly, ZnCl$^{2}$ and
OLAM are degassed at high temperature and then heated at 180 $\degree$C for 30 minutes to make a
concentrated solution of Zn$^{2}$+ for cation exchange. After cooling the Zn$^{2+}$ solution to 100 $\degree$C, an aliquot of
the solution is mixed with toluene and the temperature is adjusted to 50 $\degree$C. The synthesized roxbyite
NPs are dissolved in tri-octyl phosphine and then injected into the Zn$^{2+}$ solution in and allowed to react
for 30 minutes before quenching the reaction with cold acetone.

\textit{Cu-SiC Catalyst.} The Cu/SiC catalyst was prepared on a commercial SiC support purchased from Shanghai Yao Tian Nano Material Co., Ltd. following previously described methods \cite{li2018siCsynthesis}.  The catalyst was prepared by incipient wetness impregnation using a Cu(NO$_3$)$_2\cdot$3H$_2$O aqueous solution 0.35 gmL with 5 wt\% Cu loading followed by calcination in air at 350 $\degree$C for 2 h. 

\textit{Acrylic C-TiO$_2$ Nanoparticles.} The C-TiO$_2$ sample was prepared by blending commercial TiO$_2$ particles (purchased from Chemours) with an emulsion polymer latex. Before conducting the chemical imaging at room temperature, the blend was pre-treated under the electron beam in a Thermo Fisher T12 TEM at -80 $\degree$C to promote cross-linking in the latex and preserve its morphology above the glass transition temperature. 

\subsection*{Electron Tomography Acquisition}

Simultaneously acquired HAADF and EELS tilts series for the Au-Fe$_3$O$_4$ specimen were collected on a Talos F200X G2 (Thermo Fisher) operated at 200 keV with a probe current of 115 pA, probe semi-angle of roughly 10.5 mrad and inner collection semi-angle of 50 mrad. The HAADF projections were collected from -60$^{\circ}$ to +60$^{\circ}$ with a 3$^{\circ}$ angular increment using a Model 2021 Fischione Analytical Tomography Holder. At each tilt angle, a STEM image with a 24 $\mu$s dwell time at each pixel of a lateral dimension 6.4 Å. Simultaneously acquired HAADF and EELS spectrums were acquired at acquired with a 15$^{\circ}$ angular increment with a dwell time of 3 ms receiving a total electron dose of $4.9 \times 10^{5} ~e/$Å$^{2}$ ($1.72 \times 10^{4} ~e/$Å$^{2}$, $4.73 \times 10^{5} ~e/$Å$^{2}$ for the HAADF and EELS modality, respectively). Refer to Supplementary Fig. \ref{sfig:RawHaadfAuFeOTilts} and \ref{sfig:RawEELSAuFeOTilts} to view the raw tilt series. 


Simultaneously acquired HAADF and EELS tilt series for the Co$_3$O$_4$ - Mn$_3$O$_4$ specimen were collected on a double aberration-corrected modified FEI Titan 80-300 microscope (the TEAM I instrument at the National Center for Electron Microscopy within Lawrence Berkeley National Laboratory) operated at 300 keV with a probe current of 115 pA and semi-angle of roughly 10 mrad. This microscope is equipped with a Gatan K3 detector and Continuum spectrometer. The HAADF projections were recorded from -60$^{\circ}$ to +60$^{\circ}$ with a 3$^{\circ}$ angular increment using a Hummingbird Scientific eucentric Tomography Holder. At each tilt angle, a STEM image with a 24 $\mu$s dwell time at each pixel of a lateral dimension of 7.79 Å. Simultaneously acquired HAADF and EELS spectrums were acquired at acquired with a 15$^{\circ}$ angular increment with a dwell time of 0.677 ms receiving a total electron dose of $8.37 \times 10^{4}~e/$Å$^{2}$ ($1.16 \times 10^{4}~e/$Å$^{2}$, $7.21 \times 10^{4}~e/$Å$^{2}$ for the HAADF and EELS modality, respectively). Refer to Supplementary Fig. \ref{sfig:RawHaadfCoMnOTilts} and \ref{sfig:RawEELSCoMnOTilts} to view the raw tilt series. 

Simultaneously acquired HAADF and EDX tilt series for the Cu-SiC specimen were collected on a Talos F200X G2 (Thermo Fisher) operated at 200 keV with a probe current of 250 pA, probe semi-angle of roughly 10.5 mrad and collection angle of 44-200 mrad. The HAADF projections were collected from -75 to +70 with a 3$^{\circ}$ angular increment. At each tilt angle, a STEM image with a 20 $\mu$s at each pixel of the lateral dimension of 1.4679 nm. Simultaneously acquired HAADF and EDX spectrums were acquired at acquired with a 15$^{\circ}$ angular increment with a dwell time of 20 $\micro$s dwell time for 25 frames receiving a total electron dose of $4.33 \times 10^{4}~e/$Å$^{2}$ ($7.1 \times 10^{3}~e/$Å$^{2}$, $3.62 \times 10^{4}~e/$Å$^{2}$ for the HAADF and EELS modality, respectively). The initial chemical distributions were generated from EDX maps using commercial Velox software that produced initial net count estimates (however atomic percent estimates are also suitable).

\subsection*{Multi-Modal Tilt Series Alignment}

The EELS signals were obtained by integration over the core loss edges, all of which were done after background subtraction. The background EELS spectra were modeled using a linear combination of power laws implemented using the open source Cornell Spectrum Imager software \cite{cueva2012csi}. 

Before tilt series alignment, the spectrum images have been drift-corrected after acquisition assuming a time-dependent linear drift model, as illustrated in Supplementary Fig. \ref{sfig:DriftAlign}. The survey image, which is taken with an identical dwell time as the HAADF tilts, is taken as a reference. Iterative image registration between the chemical and HAADF signals seek an optimal translation and affine transformation. Following registration, the background of each projection was removed. For this purpose, the mean grey level in the outer regions was calculated for each projection and subtracted. In this way, the signal contribution of the carbon film could be eliminated. 

For the alignment of the tilt series, a coarse alignment is performed with either the center of mass~(CoM) or cross-correlation  method~\cite{frank1992crossCorrelation}. CoM works best when the total projected volume is fixed across specimen tilts (i.e., the object is isolated)~\cite{sanders2017l1Reg}. In cases where either of these requirements are not met (e.g. fields of view where multiple particles are visible as demonstrated with the Au - Fe$_3$O$_4$ nanoparticles), cross-correlation should be considered. Fine alignment is performed with custom written projection matching method~\cite{odstrcil2019alignment} on the HAADF modality. The measured translation shifts are subsequently applied to the corresponding tilts where simultaneously acquired chemical maps were acquired. 

\subsection*{Fused Multi-Modal Tomography Recovery}

Here, fused multi-modal electron microscopy is framed as an inverse problem expressed in the following form: $\hat{\bm{x}}~=~\arg\min_{\bm{x} \geq 0} \lambda_1 \Psi_1(\bm{x}) + \lambda_2 \Psi_2(\bm{x}) + \lambda_3 \mathrm{TV}(\bm{x})$
where $\hat{\bm{x}}$ is the final reconstruction, and the three terms are described in the main manuscript (Eq.~\ref{eq:costFunc}). When implementing an algorithm to solve this problem, we concatenate the multi-element spectral variable ($\bm{x}$) as 2D matrices: $\bm{x} \in~\mathbb{R}^{n_y \cdot n_y \cdot n_{i} \times n_x }$ where $n_{i}$ denotes the total number of reconstructed elements and $n_x, n_y$ represent number of pixels in the x and y direction and $\bm{x}_i, \bm{b}_i$ are the reconstructions and chemical maps for element $i$ $\big( \bm{x}_i \in~\mathbb{R}^{n_y \cdot n_y \times n_x }$ and $\bm{b}_i \in~\mathbb{R}^{n_y \cdot N^{\mathrm{proj}}_{\mathrm{chem}} \times n_x} \big)$. Here the axis of rotation is along the $x$-direction ($n_x$). 

The optimization problem is solved by a combination of gradient descent with total variation regularization. We minimize this cost function by iteratively descending along the negative gradient directions for the first two terms and subsequently evaluate the isotropic TV proximal operator to denoise the chemical volumes~\cite{beck2009tv}. The gradients of the first two terms are:
\begin{align}
\nabla_{\bm{x}} \Psi_1(\bm{x}) &= - \gamma \text{diag} \big(\bm{x}^{\gamma-1}\big) \bm{\Sigma}^T \bm{A}_h^{T} \Big(\bm{A}_h (\bm{\Sigma} \bm{x}^{\gamma} )^{T} - \bm{b}_{H} \Big) \\
\nabla_{\bm{x}_i} \Psi_2(\bm{x}_i) &= \bm{A}_c^T \Big( (\bm{A}_c \bm{x}_i - \bm{b}_i) \oslash (\bm{A}_c \bm{x}_i + \varepsilon) \Big),
\end{align}
where $\oslash$ denotes point-wise division, $\bm{b}_H \in \mathbb{R}^{ n_y N^{\mathrm{proj}}_{\mathrm{HAADF}} \times n_x }$ are the HAADF measurements, $\bm{A}_h \in \mathbb{R}^{n_y \cdot N_{\mathrm{HAADF}}^{\mathrm{proj}} \times n_y \cdot n_y}$ and $\bm{A}_c~\in~\mathbb{R}^{n_y \cdot N_{\mathrm{chem}}^{\mathrm{proj}} \times n_y \cdot n_y }$ are forward projection matrices operating on the chemical and HAADF modalities. Here, the first term in the cost function, relating the elastic and inelastic modalities, has been equivalently re-written as $\Psi_1~=~\frac{1}{2} \big\| \bm{A}_h (\bm{\Sigma} \bm{x}^{\gamma} )~-~ \bm{b}_{H} \big \|_2^2$, where $\bm{\Sigma} \in \mathbb{R}^{n_y \cdot n_y \times n_y \cdot n_y \cdot n_i}$ and $\bm{\Sigma} \bm{x}$ expresses the summation of all chemistries as matrix-vector multiplication. Evaluating the TV proximal operator is in itself another iterative algorithm. In addition, we impose a non-negativity constraint since negative concentrations are unrealistic. We initialize the first iterate with reconstructions composed purely of the raw measured data ($\bm{x}^0_i = \arg \min \Psi_2$). This is an ideal starting point as it is a local minimizer of $\Psi_2$. Smooth and asymptotic decay of all three terms in Eq.~\ref{eq:costFunc} is an indicator of reliable reconstruction. The final 3D HAADF and multi-modal chemical volumes were rendered using the Tomviz platform (tomviz.org~\cite{schwartz2022realTimeTomo} ). 


\subsection*{Multi-Modal Simulations and Bayesian Hyperparameter Optimization}

To demonstrate the functionality of our fused multi-modal electron tomography algorithm, we created a multi-channel phantom specimen inspired from an experimental system. The phantom consists of four channels, which we attribute to the crystal stoichiometry of CuO, CoO, and Au (Fig. \ref{fig::tomo_sim}c) with a volume size of 256$^3$. The HAADF intensity is proportional to  $\sum_e (Z_i x_i)^\gamma$ where $x_i$ reflects the element's stoichiometry. To produce chemical maps with realistic noise characteristics, we set the background (vacuum) to roughly 15$\%$ of the max intensity and subsequently applied Poisson noise to meet the desired SNR. For a Poisson-limited signal, each synthetic image has an SNR of $\frac{\mu_s + \mu_s^2}{\sigma^2_N}$  where $\mu_s$ is the mean signal and $\sigma_N^2$ is the variance of noise \cite{reed2021throughFocalSim} In the case of Figure \ref{fig::tomo_sim}, the SNR of the Co, Cu, O, Au, and HAADF modalities were 1.92, 2.89, 2.69, 1.96, 2208.67, respectively. Prior to measuring the NRMSE of the reconstructed volumes, the chemical distributions were normalized with zero mean and unit standard deviation. The NRMSE expresses a normalized measure of agreement between the reconstructed ($\bm{x}$) and ground truth ($\bm{y}$) : $ \sqrt{ \frac{\sum_{i,j,k} ( \bm{y}_{i,j,k} - \bm{x}_{i,j,k})^2}{\sum_{i,j,k} (\bm{y}_{i,j,k})^2 }}$. While the HAADF SNR may be high, we found the NRMSE reliably converges when above 50 (Supplementary Fig. \ref{sfig:chemSNR}).

Determining optimal regularization parameters for the phase diagram (Fig \ref{fig::tomo_sim}a) is computationally expensive to explore due to its variability across sampling conditions. While grid search could find the best parameters by exhaustively exploring all possible candidate values, the computation time would be expensive as each map would take approximately 125 days to complete on a single GPU. 

We efficiently explored the parameter space with Bayesian optimization (BO) --- a machine learning framework known for optimizing expensive unknown objective functions with minimal evaluations \cite{zhang2020boMatDesign, cao2022ePtychoBO}. It works by building a probabilistic model of the objective function with Gaussian processes (GP) regression. GP not only estimates our function of interest but also provides the uncertainty measurements to guide future predictions. BO takes into account past evaluations when determining future hyperparameter selections via an acquisition function \cite{mockus1994bayes}. For our simulations, we carried out BO with GP in Python with the Scikit Optimize library (scikit-optimize.github.io/stable) with the Matern kernel and GP Hedge acquisition strategy \cite{hoffman2011gpHedge}. By exploiting BO with GP, we are able to provide an atlas of balanced hyperparameters for Eq. \ref{eq:costFunc} with the CoNiO and CoCuO synthetic datasets (Supplementary Figs. \ref{sfig:CuCoParamMap}-\ref{sfig:CoNiOParamMap}). The estimated parameter landscape is smooth and continuous with a clear global optimum.


Asynchronous parallel BO on supercomputing resources allowed us to efficiently run several reconstructions simultaneously on a single node. This form of parallel computing resulted in several factors of computational speed up as multiple GPUs received unique experimental parameters (e.g. SNR or sampling) to reconstruct concurrently amongst each other. Specifically, the computation time to generate an NRMSE map was reduced by 99.8\% -- taking less than a day to complete (18 hours). In total, 3,452 GPU hours were used to complete these simulations -- 1078 hours on Summit - OLCF and 1078 hours on ThetaGPU - ALCF for the phase diagrams (Fig. \ref{fig::tomo_sim} and Supplementary Fig. \ref{sfig:CoNiOphaseMap}). An additional 1,296 GPU hours on Summit were used to produce the SNR plots (Supplementary Fig. \ref{sfig:chemSNR}).


\section*{Code Availability}
All of the multi-modal electron tomography reconstruction and iterative alignment codes are available at \href{https://github.com/jtschwar/tomo_TV}{github.com/jtschwar/tomo$\_$TV} and \href{https://github.com/jtschwar/projection_refinement}{github.com/jtschwar/projection$\_$refinement}. A sample jupyter notebook outlining the fused multi-modal reconstruction on the Cu-SiC and Au-Fe$_3$O$_4$ material systems will be available in the tomo TV repository.

\section*{Data Availability}
The raw and aligned Au-Fe$_3$O$_4$, Co$_3$O$_4$-Mn$_3$O$_4$, and Cu-SiC tilt series will be available in a Zenodo repository.